\documentclass[12pt]{article}
\usepackage{amsmath,amssymb,amsthm,amsxtra,overpic,bbm,bm,epsfig}
\def\thefootnote{\fnsymbol{footnote}}

\begin{document}

\vspace{0.2cm}

\begin{center}
{\large\bf Leptonic Unitarity Triangles and Effective Mass Triangles of \\
the Majorana Neutrinos}
\end{center}

\vspace{0.1cm}

\begin{center}
{\bf Zhi-zhong Xing}$^{1), 2), 3)}$ \footnote{E-mail:
xingzz@ihep.ac.cn} ~and~ {\bf Jing-yu Zhu}$^{1)}$
\footnote{E-mail: zhujingyu@ihep.ac.cn} \\
{\small 1) Institute of High Energy Physics, Chinese Academy of
Sciences, Beijing 100049, China \\
2) School of Physical Sciences, University of Chinese Academy of
Sciences, Beijing 100049, China \\
3) Center for High Energy Physics, Peking University, Beijing
100080, China }
\end{center}

\vspace{1.5cm}

\begin{abstract}
Given the best-fit results of six neutrino oscillation parameters,
we plot the Dirac and Majorana unitarity triangles (UTs) of the
$3\times 3$ lepton flavor mixing matrix to show their real shapes in
the complex plane. The connections of the three Majorana UTs with
neutrino-antineutrino oscillations and neutrino decays are explored,
and the possibilities of right or isosceles UTs are discussed. In
the neutrino mass limit of $m^{}_1 \to 0$ or $m^{}_3 \to 0$, which
is definitely allowed by current data, we show how the six triangles
formed by the effective Majorana neutrino masses $\langle
m\rangle^{}_{\alpha\beta}$ (for $\alpha, \beta = e, \mu, \tau$) and
their corresponding component vectors look like in the complex
plane. The relations of such triangles to the Majorana phases and to
the lepton-number-violating decays $H^{++} \to \alpha^+ \beta^+$ in
the type-II seesaw mechanism are also illustrated.
\end{abstract}

\begin{flushleft}
\hspace{0.8cm} PACS number(s): 14.60.Pq, 13.10.+q, 25.30.Pt \\
\hspace{0.8cm} Keywords: lepton flavor mixing, unitarity triangle,
Majorana mass, CP violation
\end{flushleft}

\def\thefootnote{\arabic{footnote}}
\setcounter{footnote}{0}

\newpage

\section{Introduction}

In the quark sector the language of the unitarity triangles (UTs)
has proved to be quite useful in describing weak CP violation which
is governed by the nontrivial phase of the $3\times 3$
Cabibbo-Kobayashi-Maskawa (CKM) quark flavor mixing matrix
\cite{CKM}. The same UT language was first applied to the lepton
sector in 1999 \cite{FX00} to illustrate CP violation in neutrino
oscillations, and a peculiar role of the Majorana phases in such
leptonic UTs was emphasized in 2000 \cite{Branco}. Since then a lot
of attention has been paid to this kind of geometrical description
of lepton flavor mixing and its applications in neutrino
phenomenology \cite{UT1}---\cite{He}.

Thanks to a number of well-established neutrino oscillation
experiments \cite{Wang}, one has determined the neutrino oscillation
parameters $\Delta m^2_{21}$, $|\Delta m^2_{31}|$, $\theta^{}_{12}$,
$\theta^{}_{13}$ and $\theta^{}_{23}$ to a good degree of accuracy
in the standard three-flavor scheme \cite{FIT,GG}. Although the sign
of $\Delta m^2_{31}$ remains unknown, a preliminary hint for $\delta
\sim 3\pi/2$ has been seen by combining the latest T2K \cite{T2K}
and Daya Bay \cite{DYB} data \cite{Lisi}. This progress is
remarkable, because it allows us to plot the UTs of the $3\times 3$
Pontecorvo-Maki-Nakagawa-Sakata (PMNS) lepton flavor mixing matrix
$U$ \cite{PMNS} in the complex plane to show their real shapes. One
of the purposes of the present paper is just to do this job. We are
going to classify the six UTs into two categories: three Dirac
triangles governed by the orthogonality relations
\begin{eqnarray}
\triangle^{}_e : & \hspace{0.1cm} & U^{}_{\mu 1} U^*_{\tau 1} +
U^{}_{\mu 2} U^*_{\tau 2} + U^{}_{\mu 3} U^*_{\tau 3} = 0 \; ,
\nonumber \\
\triangle^{}_\mu : & \hspace{0.1cm} & U^{}_{\tau 1} U^*_{e 1} +
U^{}_{\tau 2} U^*_{e 2} + U^{}_{\tau 3} U^*_{e 3} = 0 \; ,
\nonumber \\
\triangle^{}_\tau : & \hspace{0.1cm} & U^{}_{e 1} U^*_{\mu 1} +
U^{}_{e 2} U^*_{\mu 2} + U^{}_{e 3} U^*_{\mu 3} = 0 \; ,
\end{eqnarray}
which are insensitive to the Majorana phases of $U$; and three
Majorana triangles dictated by the orthogonality relations
\begin{eqnarray}
\triangle^{}_1 : & \hspace{0.1cm} & U^{}_{e 2} U^*_{e 3} + U^{}_{\mu
2} U^*_{\mu 3} + U^{}_{\tau 2} U^*_{\tau 3} = 0 \; ,
\nonumber \\
\triangle^{}_2 : & \hspace{0.1cm} & U^{}_{e 3} U^*_{e 1} + U^{}_{\mu
3} U^*_{\mu 1} + U^{}_{\tau 3} U^*_{\tau 1} = 0 \; ,
\nonumber \\
\triangle^{}_3 : & \hspace{0.1cm} & U^{}_{e 1} U^*_{e 2} + U^{}_{\mu
1} U^*_{\mu 2} + U^{}_{\tau 1} U^*_{\tau 2} = 0 \; ,
\end{eqnarray}
whose orientations are fixed by the Majorana phases of $U$. In
section 2 the real shapes of these six triangles will be shown with
the help of the best-fit results of six neutrino oscillation
parameters, and their uncertainties associated with the $1\sigma$
uncertainties of the input parameters will be briefly illustrated.
Furthermore, the possibilities of right or isosceles UTs in a given
neutrino mass ordering will be discussed, and the connections of the
Majorana UTs with neutrino-antineutrino oscillations and neutrino
decays will be explored.

On the other hand, we are curious about whether the reconstructed
elements of the effective Majorana neutrino mass matrix
\begin{eqnarray}
\langle m\rangle^{}_{\alpha\beta} \equiv m^{}_1 U^{}_{\alpha 1}
U^{}_{\beta 1} + m^{}_2 U^{}_{\alpha 2} U^{}_{\beta 2} + m^{}_3
U^{}_{\alpha 3} U^{}_{\beta 3} \;
\end{eqnarray}
can be similarly described in the complex plane. The answer is
affirmative, but this will involve the quadrangles instead of the
triangles in general \cite{YL1}. In the neutrino mass limit $m^{}_1
\to 0$ or $m^{}_3 \to 0$, which is compatible with current neutrino
oscillation data and allows one to remove one of the Majorana
phases, the relations in Eq. (3) will be simplified to describe six
triangles. Such mass triangles (MTs) are phenomenologically
interesting in the sense that they are directly related to some rare
but important lepton-number-violating (LNV) processes. The example
associated with the neutrinoless double-beta ($0\nu 2\beta$) decay
has recently been discussed in Ref. \cite{YL2}. In the present paper
we are going to show how each MT formed by $\langle
m\rangle^{}_{\alpha\beta}$ (for $\alpha, \beta = e, \mu, \tau$) and
its two component vectors in the $m^{}_1 \to 0$ or $m^{}_3 \to 0$
limit looks like. The relations of such triangles to the Majorana
phases and to the LNV decays $H^{++} \to \alpha^+ \beta^+$ in the
type-II seesaw mechanism will also be illustrated.

Let us stress that considering the neutrino mass limit $m^{}_1 \to
0$ or $m^{}_3 \to 0$ makes sense in several aspects. Experimentally,
this possibility is not in conflict with any available data.
Theoretically, it is consistent with the spirit of Occam's razor
\cite{Yanagida}, and either $m^{}_1 =0$ or $m^{}_3 =0$ can naturally
be obtained in a neutrino mass model (e.g., the minimal type-I
seesaw mechanism \cite{FGY}). Phenomenologically, verifying or
excluding this special case may help explore the true neutrino mass
spectrum. It is also appealing in cosmology because it implies that
today's cosmic neutrino background, whose typical temperature is
only about 1.9 K (i.e., about $1.6 \times 10^{-4}$ eV), may have
both relativistic and nonrelativistic components!

\section{Leptonic UTs}

The $3\times 3$ PMNS lepton flavor mixing matrix $U$ is commonly
parametrized as follows:
\begin{eqnarray}
U = \left(
\begin{matrix} c^{}_{12} c^{}_{13} & s^{}_{12} c^{}_{13} & s^{}_{13}
e^{-{\rm i} \delta} \cr -s^{}_{12} c^{}_{23} - c^{}_{12} s^{}_{13}
s^{}_{23} e^{{\rm i} \delta} & c^{}_{12} c^{}_{23} - s^{}_{12}
s^{}_{13} s^{}_{23} e^{{\rm i} \delta} & c^{}_{13} s^{}_{23} \cr
s^{}_{12} s^{}_{23} - c^{}_{12} s^{}_{13} c^{}_{23} e^{{\rm i}
\delta} & ~ -c^{}_{12} s^{}_{23} - s^{}_{12} s^{}_{13} c^{}_{23}
e^{{\rm i} \delta} ~ & c^{}_{13} c^{}_{23} \cr
\end{matrix} \right) P^{}_\nu \; ,
\end{eqnarray}
where $c^{}_{ij} \equiv \cos\theta^{}_{ij}$, $s^{}_{ij} \equiv
\sin\theta^{}_{ij}$ (for $ij = 12, 13, 23$), and $P^{}_\nu = {\rm
Diag}\left\{e^{{\rm i}\rho}, e^{{\rm i}\sigma}, 1\right\}$
containing two Majorana phases. For the sake of simplicity, here we
adopt the best-fit and $1\sigma$ results of six neutrino oscillation
parameters obtained in Ref. \cite{GG}:
\begin{itemize}
\item     Normal mass ordering (NMO) of the neutrinos:
$\theta^{}_{12} = 33.48^{+0.78^\circ}_{-0.75^\circ}$,
$\theta^{}_{13} = 8.50^{+0.20^\circ}_{-0.21^\circ}$, $\theta^{}_{23}
= 42.3^{+3.0^\circ}_{-1.6^\circ}$, $\delta =
306^{+39^\circ}_{-70^\circ}$, $\Delta m^2_{21} =
7.50^{+0.19}_{-0.17} \times 10^{-5} ~{\rm eV}^2$ and $\Delta
m^2_{31} = +2.457^{+0.047}_{-0.047} \times 10^{-3} ~{\rm eV}^2$;

\item     Inverted mass ordering (IMO) of the neutrinos:
$\theta^{}_{12} = 33.48^{+0.78^\circ}_{-0.75^\circ}$,
$\theta^{}_{13} = 8.51^{+0.20^\circ}_{-0.21^\circ}$, $\theta^{}_{23}
= 49.5^{+1.5^\circ}_{-2.2^\circ}$, $\delta =
254^{+63^\circ}_{-62^\circ}$, $\Delta m^2_{21} =
7.50^{+0.19}_{-0.17} \times 10^{-5} ~{\rm eV}^2$ and $\Delta
m^2_{32} = -2.449^{+0.048}_{-0.047} \times 10^{-3} ~{\rm eV}^2$.
\end{itemize}
At present the two Majorana phases $\rho$ and $\sigma$ are
completely unknown. Hence we typically take $\rho = 0$ and $\sigma =
\pi/4$ throughout this paper for the purpose of illustration. With
the help of the best-fit inputs we plot the three Dirac UTs defined
by Eq. (1) and the three Majorana UTs defined by Eq. (2) in Figures
1 and 2, respectively, to show their real shapes. Both the NMO and
IMO cases have been taken into account in our plotting, and the
inner angles of the six triangles are defined in a consistent way as
follows:
\begin{eqnarray}
\phi^{}_{\alpha i} \equiv \arg\left[-
\frac{U^{}_{\beta j} U^*_{\gamma j}} {U^{}_{\beta k} U^*_{\gamma
k}}\right] = \arg\left[- \frac{U^{}_{\beta j} U^*_{\beta k}}
{U^{}_{\gamma j} U^*_{\gamma k}}\right] \; ,
\end{eqnarray}
where the Greek and Latin subscripts keep their cyclic running over
$(e, \mu, \tau)$ and $(1, 2, 3)$, respectively. Some discussions and
comments are in order.
\begin{figure}[t]
\centerline{\includegraphics[width=15cm]{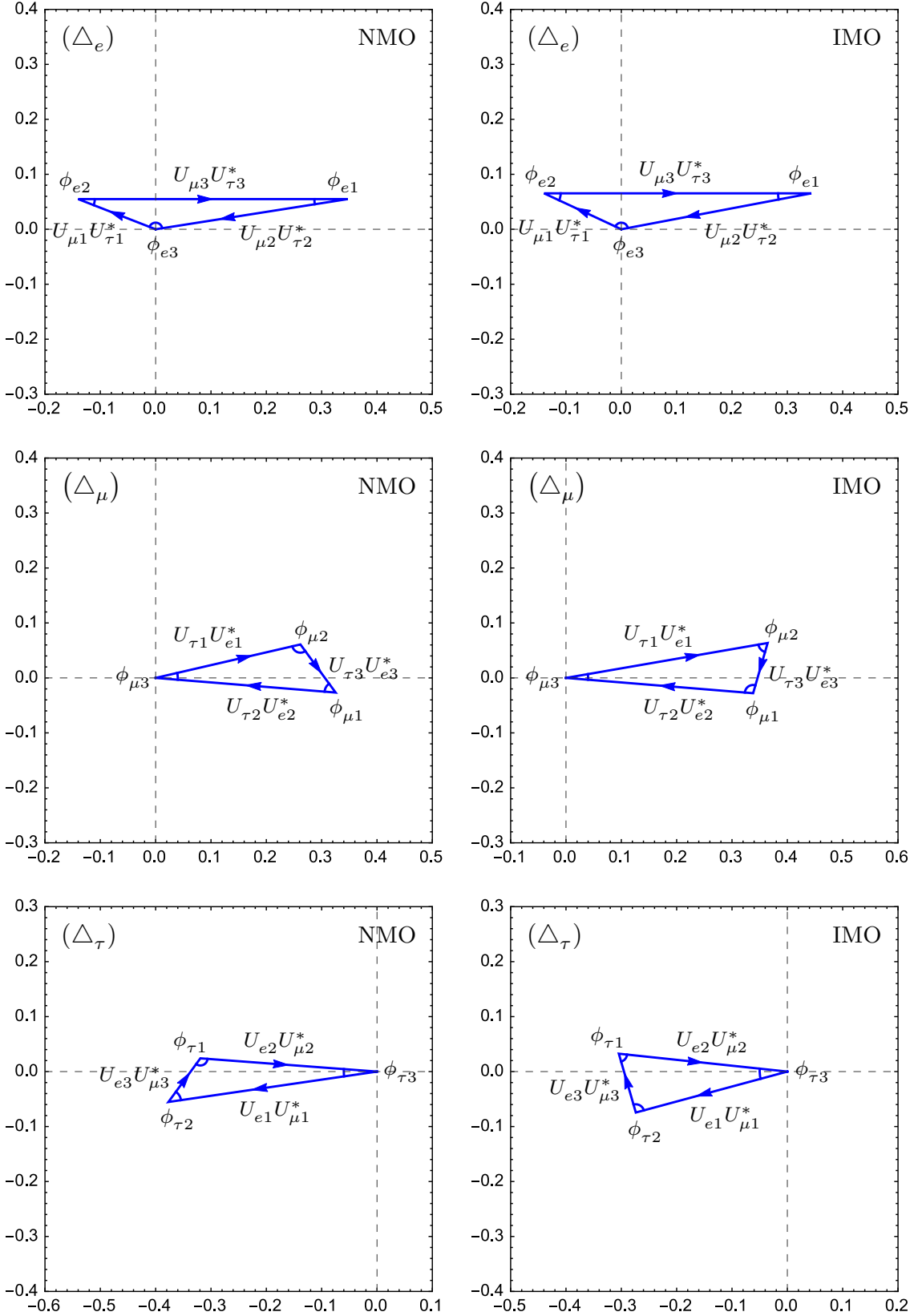}}
\caption{The real shapes of three Dirac UTs in the complex plane,
plotted by inputting the best-fit values of $\theta^{}_{12}$,
$\theta^{}_{13}$, $\theta^{}_{23}$ and $\delta$ \cite{GG} in the NMO
(left panel) or IMO (right panel) case.}
\end{figure}
\begin{figure}[t]
\centerline{\includegraphics[width=15cm]{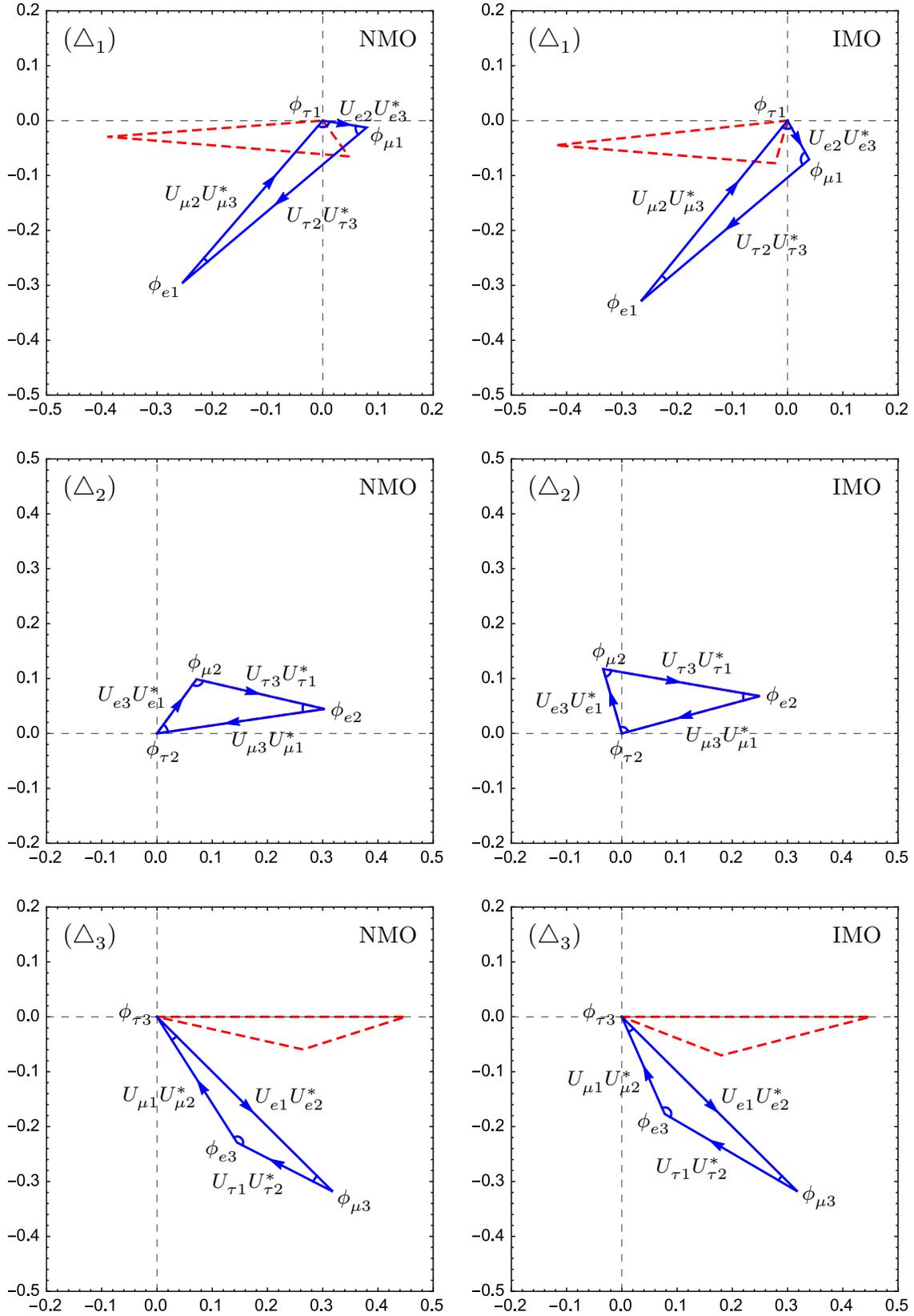}}
\caption{The real shapes and orientations of three Majorana UTs in
the complex plane, plotted by assuming the Majorana phases
$\left(\rho, \sigma\right) = \left(0, \pi/4\right)$ and inputting
the best-fit values of $\theta^{}_{12}$, $\theta^{}_{13}$,
$\theta^{}_{23}$ and $\delta$ \cite{GG} in the NMO (left panel) or
IMO (right panel) case. The dashed triangles correspond to
$\left(\rho, \sigma\right) = \left(0, 0\right)$ for comparison.}
\end{figure}

(a) In either the Dirac case or the Majorana case, the inner angles
$\phi^{}_{\mu 1}$ and $\phi^{}_{\mu 2}$ (or $\phi^{}_{\tau 1}$ and
$\phi^{}_{\tau 2}$) seem to be sensitive to the neutrino mass
ordering. To understand this, we notice
\begin{eqnarray}
\phi^{}_{\tau 1} \simeq \phi^{}_{\mu 2} \simeq \delta - \pi \; ,
\hspace{0.5cm}
\phi^{}_{\tau 2} \simeq \phi^{}_{\mu 1} \simeq 2\pi - \delta \;
\end{eqnarray}
in the leading-order approximation thanks to the relative smallness
of $\theta^{}_{13}$. Hence these four angles are actually sensitive
to the best-fit value of $\delta$, which belongs to the third
quadrant in the NMO case (i.e., $\Delta m^2_{31} >0$) or the second
quadrant in the IMO case (i.e., $\Delta m^2_{32} <0$) as given
above. In comparison, the other five inner angles of the UTs are not
so sensitive to $\delta$, and thus their results do not drastically
change in the NMO and IMO cases, as one can easily see in Figures 1
and 2.

(b) The so-called Jarlskog invariant of the $3\times 3$
PMNS matrix $U$ \cite{J} is now given as
\begin{eqnarray}
{\cal J} = \sin 2\theta^{}_{12} \cos\theta^{}_{13} \sin
2\theta^{}_{13} \sin 2\theta^{}_{23} \frac{\sin\delta}{8} \simeq
\displaystyle \left\{
\begin{array}{l}
-2.68 \times 10^{-2} \hspace{0.5cm} ({\rm NMO}) \; , \\
-3.16 \times 10^{-2} \hspace{0.5cm} ({\rm IMO}) \; ,
\end{array}
\right.
\end{eqnarray}
and it measures the strength of CP violation in neutrino
oscillations. In particular, all the areas of the six different UTs
are equal to $|{\cal J}|/2$ \cite{FX00}. On the other hand, the nine
inner angles of these triangles may form the following {\it angle}
matrix:
\begin{eqnarray}
\Phi = \begin{pmatrix} \phi^{}_{e1} & \phi^{}_{e2} & \phi^{}_{e3}
\cr \phi^{}_{\mu 1} & \phi^{}_{\mu 2} & \phi^{}_{\mu 3} \cr
\phi^{}_{\tau 1} & \phi^{}_{\tau 2} & \phi^{}_{\tau 3}
\end{pmatrix}
\simeq \displaystyle \left\{
\begin{array}{l}
\begin{pmatrix}
9.00^{\circ} & 21.44^{\circ} & 149.56^{\circ} \cr 49.33^{\circ} &
112.90^{\circ} & 17.77^{\circ} \cr 121.67^{\circ} & 45.66^{\circ} &
12.67^{\circ} \end{pmatrix}
\hspace{0.5cm} ({\rm NMO}) \; , \\
\begin{pmatrix}
10.79^{\circ} & \hspace{0.08cm} 25.11^{\circ} \hspace{0.08cm} &
144.10^{\circ} \cr 101.31^{\circ} & 64.11^{\circ} & 14.58^{\circ}
\cr 67.90^{\circ} & 90.78^{\circ} & 21.32^{\circ} \end{pmatrix}
\hspace{0.5cm} ({\rm IMO}) \; ,
\end{array}
\right.
\end{eqnarray}
whose elements satisfy the sum rules $\phi^{}_{\alpha 1} +
\phi^{}_{\alpha 2} + \phi^{}_{\alpha 3} = \phi^{}_{e i} +
\phi^{}_{\mu i} + \phi^{}_{\tau i} = \pi$ (for $\alpha = e, \mu,
\tau$ and $i = 1, 2, 3$) \cite{Luo,LuoXing}. We find that the two
off-diagonal asymmetries of $\Phi$ about its
$\phi^{}_{e1}$--$\phi^{}_{\mu 2}$--$\phi^{}_{\tau 3}$ and
$\phi^{}_{e3}$--$\phi^{}_{\mu 2}$--$\phi^{}_{\tau 1}$ axes read as
\begin{eqnarray}
{\cal A}^{}_{\rm L} \hspace{-0.2cm} & \equiv & \hspace{-0.2cm}
\phi^{}_{e2} - \phi^{}_{\mu 1} = \phi^{}_{\mu 3} - \phi^{}_{\tau 2}
= \phi^{}_{\tau 1} - \phi^{}_{e3} \simeq \left\{
\begin{array}{l}
-27.89^\circ \hspace{0.5cm} ({\rm NMO}) \; , \\ -76.20^\circ
\hspace{0.5cm} ({\rm IMO}) \; ;
\end{array}
\right.
\nonumber \\
{\cal A}^{}_{\rm R} \hspace{-0.2cm} & \equiv & \hspace{-0.2cm}
\phi^{}_{e2} - \phi^{}_{\mu 3} = \phi^{}_{\mu 1} - \phi^{}_{\tau 2}
= \phi^{}_{\tau 3} - \phi^{}_{e1} \simeq \left\{
\begin{array}{l}
+3.67^\circ \hspace{0.7cm} ({\rm NMO}) \; , \\
+10.53^\circ \hspace{0.5cm} ({\rm IMO}) \; .
\end{array}
\right.
\end{eqnarray}
These results mean that $\Phi$ only contains four independent
elements, which can reversely be used to determine the Dirac
CP-violating phase and three flavor mixing angles.

(c) The three Majorana UTs are more interesting in the sense that
their orientations depend on the values of the phase parameters
$\rho$ and $\sigma$. Even though the UTs collapsed into lines in the
$\delta = 0$ (or $\pi$) case, there would exist leptonic CP
violation unless those lines happened to lie in the abscissa or
ordinate axis. This point was first observed in Ref. \cite{Branco},
and it has been clearly illustrated in Figure 2 with the typical
inputs $\rho =0$ and $\sigma =\pi/4$. In fact, the orientations of
triangles $\triangle^{}_1$, $\triangle^{}_2$ and $\triangle^{}_3$
depend respectively on $\sigma$, $-\rho$ and $\rho-\sigma$ in the
chosen parametrization of $U$. That is why $\triangle^{}_2$ keeps
unchanged when the $(\rho, \sigma) = (0, 0)$ case is shifted to the
$(\rho, \sigma) = (0, \pi/4)$ case in our plotting. In general, the
Majorana UTs $\triangle^{}_1$, $\triangle^{}_2$ and $\triangle^{}_3$
with arbitrary values of $\rho$ and $\sigma$ can be obtained through
rotating their counterparts with $\rho =\sigma =0$ anticlockwise by
$\sigma$, $-\rho$ and $\rho -\sigma$, respectively.

(d) To reflect the Majorana nature of the PMNS matrix $U$, one may
redefine the Majorana phases as follows: $\psi^{}_{\alpha i} \equiv
\arg\left( U^{}_{\alpha j} U^*_{\alpha k}\right)$ with the Latin
subscripts running over $(1, 2, 3)$ in a cyclic way. These phases
are independent of the phases of three charged leptons, and they
form the following {\it phase} matrix:
\begin{eqnarray}
\Psi = \begin{pmatrix} \psi^{}_{e1} & \psi^{}_{e2} &
\psi^{}_{e3} \cr \psi^{}_{\mu 1} & \psi^{}_{\mu 2} & \psi^{}_{\mu 3}
\cr \psi^{}_{\tau 1} & \psi^{}_{\tau 2} & \psi^{}_{\tau 3}
\end{pmatrix}
\simeq \displaystyle \left\{
\begin{array}{l}
\begin{pmatrix}
-9.00^{\circ} & 54.00^{\circ} & -45.00^{\circ} \cr 49.33^{\circ} &
-171.66^{\circ} & 122.33^{\circ} \cr -139.67^{\circ} &
-13.10^{\circ} & 152.77^{\circ} \end{pmatrix}
\hspace{0.5cm} ({\rm NMO}) \; , \\
\begin{pmatrix}
-61.00^{\circ} & \hspace{0.08cm} 106.00^{\circ} \hspace{0.08cm} &
-45.00^{\circ} \cr 51.10^{\circ} & -164.78^{\circ} & 113.68^{\circ}
\cr -139.69^{\circ} & -9.89^{\circ} & 149.58^{\circ} \end{pmatrix}
\hspace{0.5cm} ({\rm IMO}) \; ,
\end{array}
\right.
\end{eqnarray}
where we have used the same inputs as those in obtaining Eq. (8),
and taken $\rho =0$ and $\sigma =\pi/4$ for illustration. It is
obvious that the nine elements in the three rows of $\Psi$ satisfy
the sum rules $\psi^{}_{\alpha 1} + \psi^{}_{\alpha 2} +
\psi^{}_{\alpha 3} = 0$ (for $\alpha = e, \mu, \tau$) \cite{Luo},
but those in the three columns do not have a definite correlation.
Hence the number of independent parameters in $\Psi$ is six, two
more as compared with that in $\Phi$. Given Eq. (5), the nine inner
angles of the sixe UTs can be expressed in terms of the nine
elements of $\Psi$ as
\begin{eqnarray}
\phi^{}_{\alpha i} = \psi^{}_{\beta i} - \psi^{}_{\gamma i} \pm
\pi \; ,
\end{eqnarray}
in which the Greek subscripts run over $(e, \mu, \tau)$ cyclically,
and the ``$\pm$" sign should be taken in a proper way to assure
$\phi^{}_{\alpha i} \in \left[0, \pi\right)$.

The numerical results for the shapes and inner angles of
$\triangle^{}_\alpha$ (for $\alpha = e, \mu, \tau$) and
$\triangle^{}_i$ (for $i =1, 2, 3$) given above are subject to the
inputs of the best-fit values of $\theta^{}_{12}$, $\theta^{}_{13}$,
$\theta^{}_{23}$ and $\delta$. Among them, $\delta$ involves the
largest uncertainty. One may fix the values of the three flavor
mixing angles to check how the shapes of six UTs change with
different values of $\delta$. Instead of making such a check by
taking some numerical examples, let us outline a general observation
based on the leading-order analytical approximations made in Eq.
(6). It is clear how $\phi^{}_{\tau 1} \simeq \phi^{}_{\mu 2} \simeq
\delta - \pi$ and $\phi^{}_{\tau 2} \simeq \phi^{}_{\mu 1} \simeq
2\pi - \delta$ vary with the change of $\delta$. Because
$\phi^{}_{\mu 1} + \phi^{}_{\mu 2} \simeq \phi^{}_{\mu 1} +
\phi^{}_{\tau 1} \simeq \phi^{}_{\tau 1} + \phi^{}_{\tau 2} \simeq
\phi^{}_{\mu 2} + \phi^{}_{\tau 2} \simeq \pi$ holds as a
consequence of the above approximations, the inner angles
$\phi^{}_{e 1}$, $\phi^{}_{e 2}$, $\phi^{}_{\mu 3}$ and
$\phi^{}_{\tau 3}$ must be small as required by the unitarity
conditions, and hence $\phi^{}_{e 3}$ must be the largest inner
angle. This general analytical observation is actually supported by
the explicit numerical results shown in Eq. (8).

When the uncertainties of all the four input parameters are taken
into account, the situation will become quite messy. To illustrate,
let us consider the $1\sigma$ intervals of the input quantities and
calculate the nine inner angles of six UTs. As illustrated in Table
1, the $1\sigma$ uncertainty of each inner angle is rather
significant as compared with its best-fit outcome, implying a
remarkable change of the shape of each UT. A direct illustration of
such uncertainties of $\triangle^{}_\alpha$ and $\triangle^{}_i$ in
the complex plane is difficult, since all the sides and inner angles
will deviate from those in the best-fit case (i.e., in Figures 1 and
2). At present one possible way out is to rescale and rotate each UT
to make two of its three vertices always locate at the $(0, 0)$ and
$(1, 0)$ points in the horizontal coordinate axis \cite{GG}. In this
case, however, the uncertainty associated with the third vertex of
each rescaled UT remains quite significant. Although the sides of
each real UT and those of its rescaled counterpart are different,
the inner angles of these two triangles are exactly the same. So
Table 1 is almost equally helpful for illustrating how the shapes of
six UTs are sensitive to the uncertainties of $\theta^{}_{12}$,
$\theta^{}_{13}$, $\theta^{}_{23}$ and especially $\delta$. Once
$\delta$ is determined in the next-generation accelerator-based
neutrino oscillation experiments, it will be possible for us to see
the true shapes of leptonic UTs to a reasonably good degree of
accuracy, just as we have seen the true shapes of six CKM UTs in the
quark sector today
\footnote{It is worth pointing out that the area of each CKM UT is
equal to ${\cal J}^{}_q /2 \simeq 1.53 \times 10^{-5}$ \cite{PDG},
where ${\cal J}^{}_q \simeq 3.06 \times 10^{-5}$ is the Jarlskog
invariant of the CKM quark flavor mixing matrix. This result is
about three orders of magnitude smaller than the area of each
leptonic UT shown in Figure 1 or 2, where $\delta \simeq 306^\circ$
(NMO) or $254^\circ$ (IMO) has been typically input.}.
\begin{table}[t]
\centering \caption{The numerical results for nine inner angles of
the six UTs obtained with the inputs of the best-fit values and
$1\sigma$ ranges of $\theta^{}_{12}$, $\theta^{}_{13}$,
$\theta^{}_{23}$ and $\delta$ \cite{GG}. Note that the unitarity
conditions $\phi^{}_{\alpha 1} + \phi^{}_{\alpha 2} +
\phi^{}_{\alpha 3} = \phi^{}_{e i} + \phi^{}_{\mu i} + \phi^{}_{\tau
i} = \pi$ must hold (for $\alpha = e, \mu, \tau$ and $i = 1, 2,
3$).} \vspace{0.2cm}
\begin{tabular}{ccccccccc}\hline\hline
&& \multicolumn{3}{c}{Normal mass ordering (NMO)} &&
\multicolumn{3}{c}{Inverted mass ordering (IMO)} \\ \hline
&&best-fit & & $\pm 1\sigma$ range &&best-fit & & $\pm 1\sigma$
range\\ \hline
$\phi_{e1}^{}$ & & $9.00^\circ$ && $2.73^\circ$ --- $11.90^\circ$
&& $10.79^\circ$ && $2.19^\circ$ --- $12.03^\circ$\\
$\phi_{e 2}^{}$ && $21.44^\circ$ && $6.52^\circ$ --- $26.81^\circ$
&& $25.11^\circ$ && $5.47^\circ$ --- $27.13^\circ$\\
$\phi_{e3}^{}$ && $149.56^\circ$ && $142.00^\circ$ ---
$170.54^\circ$ && $144.10^\circ$ && $141.60^\circ$---
$172.20^\circ$\\   \hline
$\phi_{\mu 1}^{}$ && $49.33^\circ$ && $13.41^\circ$ ---
$119.42^\circ$
&& $103.31^\circ$ && $39.53^\circ$ --- $167.04^\circ$\\
$\phi_{\mu 2}^{}$ && $112.90^\circ$ && $44.87^\circ$ ---
$161.11^\circ$ && $64.11^\circ$ && $9.87
^\circ$ --- $129.39^\circ$\\
$\phi_{\mu3}^{}$ && $17.77^\circ$ && $5.21^\circ$ --- $22.03^\circ$
&& $14.58^\circ$ && $2.78^\circ$--- $17.52^\circ$\\   \hline
$\phi_{\tau1}^{}$ && $121.67^\circ$ && $51.34^\circ$ ---
$163.72^\circ$
&& $67.90^\circ$ && $10.65^\circ$ --- $132.77^\circ$\\
$\phi_{\tau 2}^{}$ && $45.66^\circ$ && $12.13^\circ$ ---
$114.46^\circ$
&& $90.78^\circ$ && $33.72^\circ$ --- $164.48^\circ$\\
$\phi_{\tau3}^{}$ && $12.67^\circ$ && $3.66^\circ$ --- $19.17^\circ$
&& $21.32^\circ$ && $4.67^\circ$--- $23.38^\circ$\\   \hline\hline
\end{tabular}
\end{table}

An interesting question that one may ask is whether one of the Dirac
or Majorana UTs can be a special triangle, such as the {\it right}
triangle or the {\it isosceles} triangle. This question makes sense
because there {\it do} exist two right UTs ($\triangle^{}_c$ and
$\triangle^{}_s$) in the quark sector \cite{Xing2009} as indicated
by current experimental data. If one is only concerned about Figures
1 and 2 plotted by inputting the best-fit values of
$\theta^{}_{12}$, $\theta^{}_{13}$, $\theta^{}_{23}$ and $\delta$,
then the Dirac triangle $\triangle^{}_\tau$ and the Majorana
triangle $\triangle^{}_2$ can be regarded as the right triangles
with $\phi^{}_{\tau 2}$ being very close to $\pi/2$ in the IMO case.
This point is also clear in Eq. (8), where $\phi^{}_{\tau 2} \simeq
90.78^\circ$ has been given. If $\phi^{}_{\tau 2} =\pi/2$ holds
exactly, then one will be able to obtain the following correlation
between the Dirac phase and three flavor mixing angles:
\begin{eqnarray}
\cos\delta = -\cot\theta^{}_{12} \tan\theta^{}_{23}
\sin\theta^{}_{13} \; ,
\end{eqnarray}
implying that $\delta$ must deviate from $3\pi/2$ (or equivalently,
$-\pi/2$) to some extent and lies in the third quadrant. Such an
interesting relation can be tested with much more accurate neutrino
oscillation data to be achieved in the foreseeable future,
especially after $\delta$ is experimentally determined or
constrained.

From a model-building point of view, the $\mu$-$\tau$ reflection
symmetry should be the simplest and most natural flavor symmetry
behind the observed pattern of neutrino mixing \cite{Shun2014,Zhao}.
It predicts $\delta = 3\pi/2$ and $\theta^{}_{23} = \pi/4$, and
therefore one is left with $|U^{}_{\mu i}| = |U^{}_{\tau i}|$ (for
$i=1,2,3$). In this special case, we find that the three Majorana
triangles $\triangle^{}_i$ turn out to be the isosceles triangles
with $\phi^{}_{\mu i} = \phi^{}_{\tau i}$ (for $i =1,2,3$). Such a
possibility is not consistent with the best-fit results of current
experimental data, as one can see in either Eq. (8) or Figure 2, but
it cannot be excluded if the $2\sigma$ or $3\sigma$ results of a
global fit is taken into account. In comparison with the Majorana
triangles, the Dirac triangles are not sensitive to the $\mu$-$\tau$
reflection symmetry.

In Ref. \cite{He} it has been pointed out that the inner angles of
three Dirac UTs can directly be related to the probabilities of
normal neutrino oscillations. Here let us establish the direct
relations between the Majorana phases $\psi^{}_{\alpha i}$ defined
above Eq. (10) and the probabilities of neutrino-antineutrino
oscillations given in Ref. \cite{ZhouYL}. The results are
\begin{eqnarray}
P\left(\nu^{}_{\alpha} \to \overline{\nu}^{}_{\beta}\right)
\hspace{-0.2cm} & \equiv & \hspace{-0.2cm} \frac{\vert
K\vert^2}{E^2} \left[\sum_{i} m_i^2 \left(S_{\gamma i}^{\rm D}
\right)^2 + 2\sum_{i<j} m^{}_i m^{}_j S_{\alpha k}^{\rm M} S_{\beta
k}^{\rm M} \cos\left(2\Delta^{}_{ji} - \psi^{}_{\alpha k} -
\psi^{}_{\beta k}\right) \right] \; ,
\nonumber \\
P\left(\overline{\nu}^{}_{\alpha} \to \nu^{}_{\beta}\right)
\hspace{-0.2cm} & \equiv & \hspace{-0.2cm} \frac{\vert \overline
K\vert^2}{E^2} \left[\sum_{i} m_i^2 \left(S_{\gamma i}^{\rm D}
\right)^2 + 2\sum_{i<j} m^{}_i m^{}_j S_{\alpha k}^{\rm M} S_{\beta
k}^{\rm M} \cos\left(2\Delta^{}_{ji} + \psi^{}_{\alpha k} +
\psi^{}_{\beta k}\right) \right] \; , \hspace{0.6cm}
\end{eqnarray}
where the subscripts $(\alpha, \beta, \gamma)$ run over $(e, \mu,
\tau)$ cyclically, $K$ and $\overline K$ are the kinematical factors
independent of the index $i$ (and they satisfy $|K| =
|\overline{K}|$), $S^{\rm D}_{\alpha i} \equiv |U^{}_{\beta i}
U^{*}_{\gamma i}|$ defines one side of the Dirac UTs, $S^{\rm
M}_{\alpha i} \equiv |U^{}_{\alpha j} U^{*}_{\alpha k}|$ defines one
side of the Majorana UTs with $(i, j, k)$ running over $(1, 2, 3)$
cyclically, and $\Delta^{}_{ji} \equiv \Delta m^2_{ji} L/\left(4
E\right)$ with $\Delta m^2_{ji} = m^2_j - m^2_i$. It is therefore
clear that the difference between the probabilities of
$\nu^{}_\alpha \to \overline{\nu}^{}_\beta$ and
$\overline{\nu}^{}_\alpha \to \nu^{}_\beta$ oscillations,
\begin{eqnarray}
P\left(\nu^{}_{\alpha} \to \overline{\nu}^{}_{\beta}\right)
- P\left(\overline{\nu}^{}_{\alpha} \to \nu^{}_{\beta}\right)
= 4 \frac{\vert K\vert^2}{E^2} \sum_{i<j} \left[m^{}_i m^{}_j
S_{\alpha k}^{\rm M} S_{\beta k}^{\rm M} \sin \left(2\Delta^{}_{ji}
\right) \sin\left(\psi^{}_{\alpha k} + \psi^{}_{\beta k}\right)
\right] \; ,
\end{eqnarray}
results from the nontrivial values of the
Majorana phases. On the other hand, the rates of $\nu^{}_i \to
\nu^{}_j + \gamma$ decays in the rest frame of $\nu^{}_i$ (for
$m^{}_i > m^{}_j$) can be expressed as
\begin{eqnarray}
\Gamma^{\rm (M)}_{\nu^{}_i \to \nu^{}_j + \gamma} =
\frac{9\alpha^{}_{\rm em} G_{\rm F}^2 m_i^2}{2^{10} \pi^4 M_W^4}
\left(1-\frac{m_j^2}{m_i^2}\right)^3 \left[\left(1 +
\frac{m_j^2}{m_i^2}\right) X -\frac{2m^{}_j}{m^{}_i} Y \right] \; ,
\end{eqnarray}
where
\begin{eqnarray}
X \hspace{-0.2cm} & \equiv & \hspace{-0.2cm} \sum_{\alpha}
m_{\alpha}^4 \left(S_{\alpha k}^{\rm M} \right)^2 - \sum_{\alpha
\neq\beta} m_{\alpha}^2 m_{\beta}^2 S_{\alpha k}^{\rm M} S_{\beta
k}^{\rm M} \cos\left(\psi^{}_{\alpha k} - \psi^{}_{\beta k}\right)
\; ,
\nonumber \\
Y \hspace{-0.2cm} & \equiv & \hspace{-0.2cm} \sum_{\alpha}
m_{\alpha}^4 \left(S_{\alpha k}^{\rm M} \right)^2
\cos\left(2\psi^{}_{\alpha k}\right) + \sum_{\alpha \neq
\beta}m_{\alpha}^2 m_{\beta}^2 S_{\alpha k}^{\rm M} S_{\beta k}^{\rm
M} \cos\left(\psi^{}_{\alpha k} + \psi^{}_{\beta k}\right) \;
\end{eqnarray}
with $\alpha$ and $\beta$ running over $e$, $\mu$ and $\tau$. Since
such decay modes are CP-conserving, their rates remain finite even
if all the Majorana phases vanish.

Although both neutrino-antineutrino oscillations and neutrino decays
are undetectable at present, Eqs. (13)---(16) show that their
sensitivities to the Majorana UTs are conceptually interesting and
thus deserve a careful study. Note that all the sides of the Dirac
UTs (i.e., $S^{\rm D}_{\alpha i}$) can be determined from the {\it
appearance} experiments of normal neutrino oscillations, and all the
sides of the Majorana UTs (i.e., $S^{\rm M}_{\alpha i}$) are
measurable in the {\it disappearance} experiments of normal neutrino
oscillations \cite{Zhu}. Hence only the absolute neutrino mass scale
and the CP-violating phases are still unknown in the probabilities
of neutrino-antineutrino oscillations and the rates of neutrino
decays shown above. If such rare processes can really be measured in
the future, it will be greatly useful for probing the Majorana
phases of massive neutrinos. In practice, the $0\nu 2\beta$ decay is
the only LNV process that is being searched for in depth at low
energies, and its effective neutrino mass can be expressed as
\footnote{The treatment in Eq. (17) is currently most reasonable in
the sense that the present data cannot rule out the possibility of
$m^{}_1 =0$ or $m^{}_3 =0$. In either of these two special but
interesting cases, one of the Majorana phases will disappear,
leading to a much simpler expression of $|\langle m\rangle^{}_{ee}|$
as one will see in section 3.}
\begin{eqnarray}
\left|\langle m\rangle^{}_{ee}\right|
\hspace{-0.2cm} & = & \hspace{-0.2cm} \displaystyle
m^{}_2 |U^{}_{e 2}|^2 \left| 1 + \frac{m^{}_1}{m^{}_2} \cdot
\frac{U^2_{e 1}}{U^2_{e 2}} + \frac{m^{}_3}{m^{}_2} \cdot
\frac{U^2_{e 3}}{U^2_{e 2}}\right|
\nonumber \\
& = & \hspace{-0.2cm} \displaystyle m^{}_2 |U^{}_{e 2}|^2
\left| 1 + \frac{m^{}_1}{m^{}_2} \left|\frac{U^{}_{e 1}}{U^{}_{e 2}}
\right|^2 e^{+2 {\rm i} \psi^{}_{e 3}} + \frac{m^{}_3}{m^{}_2}
\left|\frac{U^{}_{e 3}}{U^{}_{e 2}} \right|^2
e^{-2 {\rm i} \psi^{}_{e 1}} \right| \; .
\end{eqnarray}
So a measurement of $|\langle m\rangle^{}_{ee}|$ will allow us to
constrain $\psi^{}_{e 1}$ and $\psi^{}_{e 3}$, but more experimental
information from some other LNV processes is needed in order to
fully determine these two Majorana phases in the standard
three-flavor neutrino mixing scheme.

\section{Effective MTs}

Since the $3\times 3$ Majorana mass matrix totally involves six
independent elements defined in Eq. (3), one may extend the exercise
done in Eq. (17) to reexpress the effective Majorana mass terms
$\langle m\rangle^{}_{\alpha\beta}$ as follows:
\begin{eqnarray}
\langle m\rangle^{}_{\alpha\beta}
\hspace{-0.2cm} & = & \hspace{-0.2cm} \displaystyle
m^{}_2 U^{}_{\alpha 2} U^{}_{\beta 2} \left(1 +
\frac{m^{}_1}{m^{}_2} \cdot \frac{U^{}_{\alpha 1}
U^{}_{\beta 1}}{U^{}_{\alpha 2} U^{}_{\beta 2}}
+ \frac{m^{}_3}{m^{}_2} \cdot \frac{U^{}_{\alpha 3} U^{}_{\beta 3}}
{U^{}_{\alpha 2} U^{}_{\beta 2}} \right)
\nonumber \\
& = & \hspace{-0.2cm} \displaystyle
m^{}_2 U^{}_{\alpha 2} U^{}_{\beta 2} \left[ 1 +
\frac{m^{}_1}{m^{}_2} \left|\frac{U^{}_{\alpha 1}
U^{}_{\beta 1}}{U^{}_{\alpha 2} U^{}_{\beta 2}}\right|
e^{+{\rm i} \left(\psi^{}_{\alpha 3} + \psi^{}_{\beta 3}\right)}
+ \frac{m^{}_3}{m^{}_2} \left|\frac{U^{}_{\alpha 3} U^{}_{\beta 3}}
{U^{}_{\alpha 2} U^{}_{\beta 2}}\right|
e^{-{\rm i} \left(\psi^{}_{\alpha 1} + \psi^{}_{\beta 1}\right)}
\right] \; , \hspace{0.5cm}
\end{eqnarray}
where $\alpha$ and $\beta$ run over $e$, $\mu$ and $\tau$. In the
complex plane Eq. (18) represents six quadrangles whose inner angles
are some combinations of the Majorana phases. But such a geometrical
description is so complicated that it might not be very useful for
neutrino phenomenology. For this reason, we shall subsequently focus
on a much simpler but interesting situation.

It is obvious that one of the two phase combinations in Eq. (18) can
always be rotated away in the neutrino mass limit $m^{}_1 \to 0$ or
$m^{}_3 \to 0$. Given the phase convention of the PMNS matrix $U$ in
Eq. (4), one may simply switch off $\rho$ so as to fit the $m^{}_1
=0$ or $m^{}_3 =0$ case. For this reason, we write out the explicit
expressions of the six effective Majorana neutrino masses defined in
Eq. (3) by setting $\rho =0$:
\begin{eqnarray}
\langle m \rangle^{}_{ee} \hspace{-0.2cm} & \equiv & \hspace{-0.2cm}
\displaystyle
m^{}_1 c_{12}^2 c_{13}^2 + m^{}_2 s_{12}^2 c_{13}^2 e^{2{\rm
i}\sigma} + m^{}_3 s_{13}^2 e^{-2{\rm i} \delta} \; ,
\nonumber \\
\langle m \rangle^{}_{\mu \mu} \hspace{-0.2cm} & \equiv &
\hspace{-0.2cm} \displaystyle
m^{}_1 \left(s^{}_{12} c^{}_{23} + c^{}_{12}
s^{}_{13} s^{}_{23} e^{{\rm i}\delta} \right)^2 + m^{}_2
\left(c^{}_{12} c^{}_{23} - s^{}_{12} s^{}_{13} s^{}_{23} e^{{\rm
i}\delta}\right)^2 e^{2{\rm i}\sigma} + m^{}_3 c_{13}^2 s_{23}^2 \; ,
\nonumber \\
\langle m \rangle^{}_{\tau \tau} \hspace{-0.2cm} & \equiv &
\hspace{-0.2cm} \displaystyle
m^{}_1 \left(s^{}_{12} s^{}_{23} - c^{}_{12}
s^{}_{13} c^{}_{23} e^{{\rm i}\delta}\right)^2 + m^{}_2
\left(c^{}_{12} s^{}_{23} + s^{}_{12} s^{}_{13} c^{}_{23} e^{{\rm
i}\delta}\right)^2 e^{2{\rm i}\sigma} + m^{}_3 c_{13}^2 c_{23}^2 \; ;
\nonumber \\
\langle m \rangle^{}_{e \mu} \hspace{-0.2cm} & \equiv &
\hspace{-0.2cm} \displaystyle
-m^{}_1 c^{}_{12} c^{}_{13} \left(s^{}_{12}
c^{}_{23} + c^{}_{12} s^{}_{13} s^{}_{23} e^{{\rm i}\delta}\right) +
m^{}_2 s^{}_{12} c^{}_{13} \left(c^{}_{12} c^{}_{23} - s^{}_{12}
s^{}_{13} s^{}_{23} e^{{\rm i}\delta}\right) e^{2{\rm i}\sigma}
\nonumber \\
&& \displaystyle
+ m^{}_3 c^{}_{13} s^{}_{13} s^{}_{23} e^{-{\rm i}\delta} \; ,
\nonumber \\
\langle m \rangle^{}_{e \tau} \hspace{-0.2cm} & \equiv &
\hspace{-0.2cm} \displaystyle
m^{}_1 c^{}_{12} c^{}_{13} \left(s^{}_{12} s^{}_{23}
- c^{}_{12} s^{}_{13} c^{}_{23} e^{{\rm i}\delta}\right) - m^{}_2
s^{}_{12} c^{}_{13} \left(c^{}_{12} s^{}_{23} + s^{}_{12} s^{}_{13}
c^{}_{23} e^{{\rm i}\delta}\right) e^{2{\rm i}\sigma}
\nonumber \\
&& \displaystyle
+ m^{}_3 c^{}_{13} s^{}_{13} c^{}_{23} e^{-{\rm i}\delta} \; ,
\nonumber \\
\langle m \rangle^{}_{\mu \tau} \hspace{-0.2cm} & \equiv &
\hspace{-0.2cm} \displaystyle
-m^{}_1 \left(s^{}_{12} s^{}_{23} - c^{}_{12}
s^{}_{13} c^{}_{23} e^{{\rm i}\delta}\right) \left(c^{}_{23}
s^{}_{12} + c^{}_{12} s^{}_{13} s^{}_{23} e^{{\rm i}\delta}\right )
\nonumber \\
&& \displaystyle
-m^{}_2 \left(c^{}_{12} s^{}_{23} + s^{}_{12} s^{}_{13} c^{}_{23}
e^{{\rm i}\delta}\right) \left(c^{}_{12} c^{}_{23} - s^{}_{12}
s^{}_{13} s^{}_{23} e^{{\rm i}\delta}\right) e^{2{\rm i}\sigma} +
m^{}_3 c_{13}^2 c^{}_{23} s^{}_{23} \; . \hspace{0.4cm}
\end{eqnarray}
Then it is much easier to consider the $m^{}_1 \to 0$ or $m^{}_3 \to
0$ limit, in which $\langle m\rangle^{}_{\alpha\beta}$ and its two
component vectors form a {\it mass} triangle (MT) in the complex
plane.

In view of the best-fit values of two neutrino mass-squared
differences reported by Gonzalez-Garcia {\it et al} \cite{FIT}, we
obtain $m^{}_2 \simeq 0.0087$ eV and $m^{}_3 \simeq 0.0496$ eV in
the $m^{}_1 \to 0$ limit (NMO); or $m^{}_1 \simeq 0.0487$ eV and
$m^{}_2 \simeq 0.0495$ eV in the $m^{}_3 \to 0$ limit (IMO). In
either case one may plot the six effective MTs with the help of Eq.
(16), the best-fit values of $\theta^{}_{12}$, $\theta^{}_{13}$,
$\theta^{}_{23}$ and $\delta$, and the assumption of $\sigma
=\pi/4$. Our results about the MTs $\triangle A^{}_i B^{}_i C^{}_i$
(for $i=1, 2, \cdots, 6$, NMO) or $\triangle D^{}_i E^{}_i F^{}_i$
(for $i=1, 2, \cdots, 6$, IMO) are shown in Figures 3 and 4,
respectively. Some discussions and comments are in order.
\begin{figure}[t]
\centerline{\includegraphics[width=15cm]{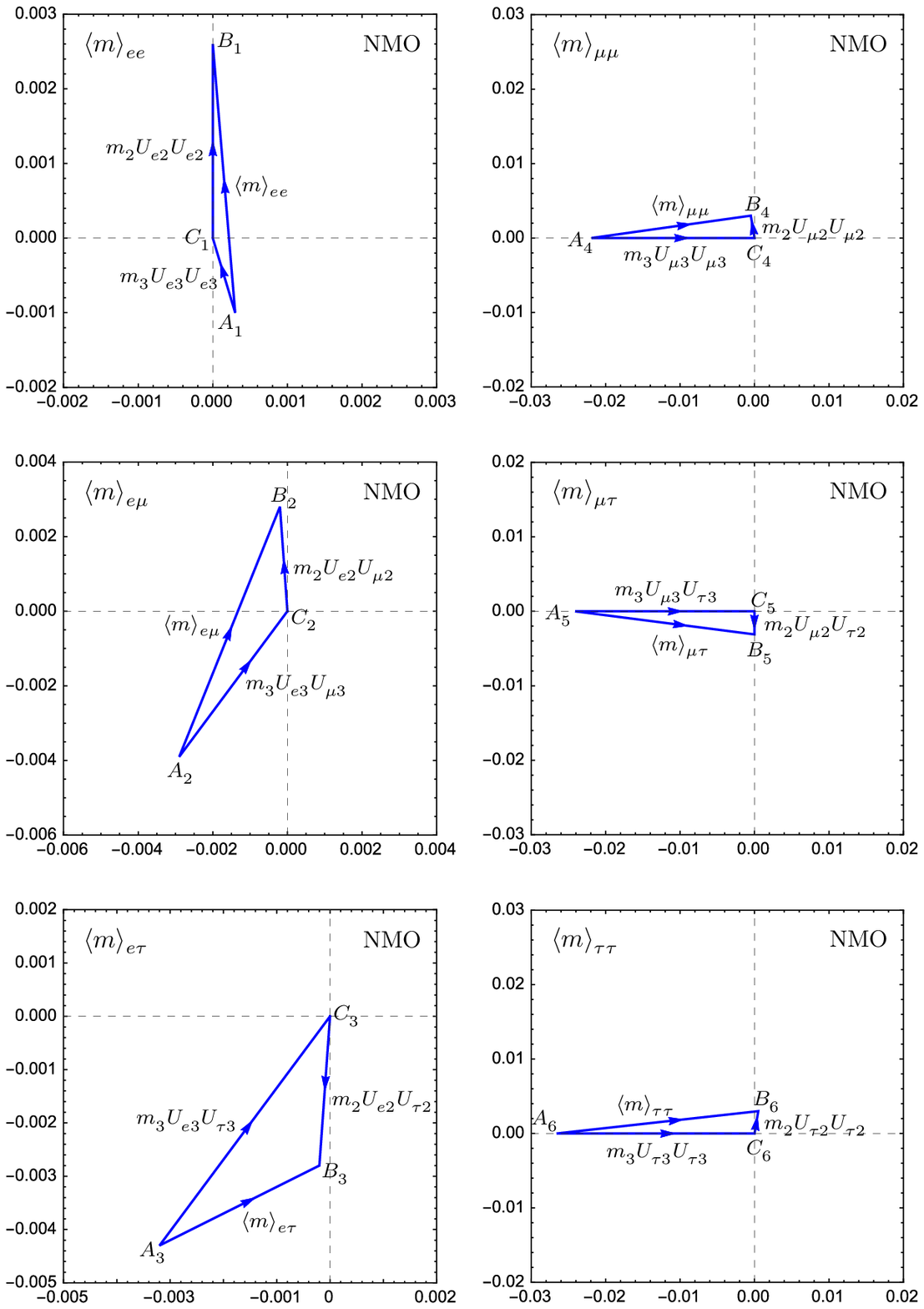}}
\caption{Six effective MTs $\triangle A^{}_i B^{}_i C^{}_i$ (for
$i=1, 2, \cdots, 6$) of the Majorana neutrinos in the $m^{}_1 \to 0$
limit in the complex plane, plotted by assuming the Majorana phase
$\sigma = \pi/4$ and inputting the best-fit values of $\Delta
m^2_{21}$, $\Delta m^2_{31}$, $\theta^{}_{12}$, $\theta^{}_{13}$,
$\theta^{}_{23}$ and $\delta$ \cite{GG} in the NMO case.}
\end{figure}
\begin{figure}[t]
\centerline{\includegraphics[width=15cm]{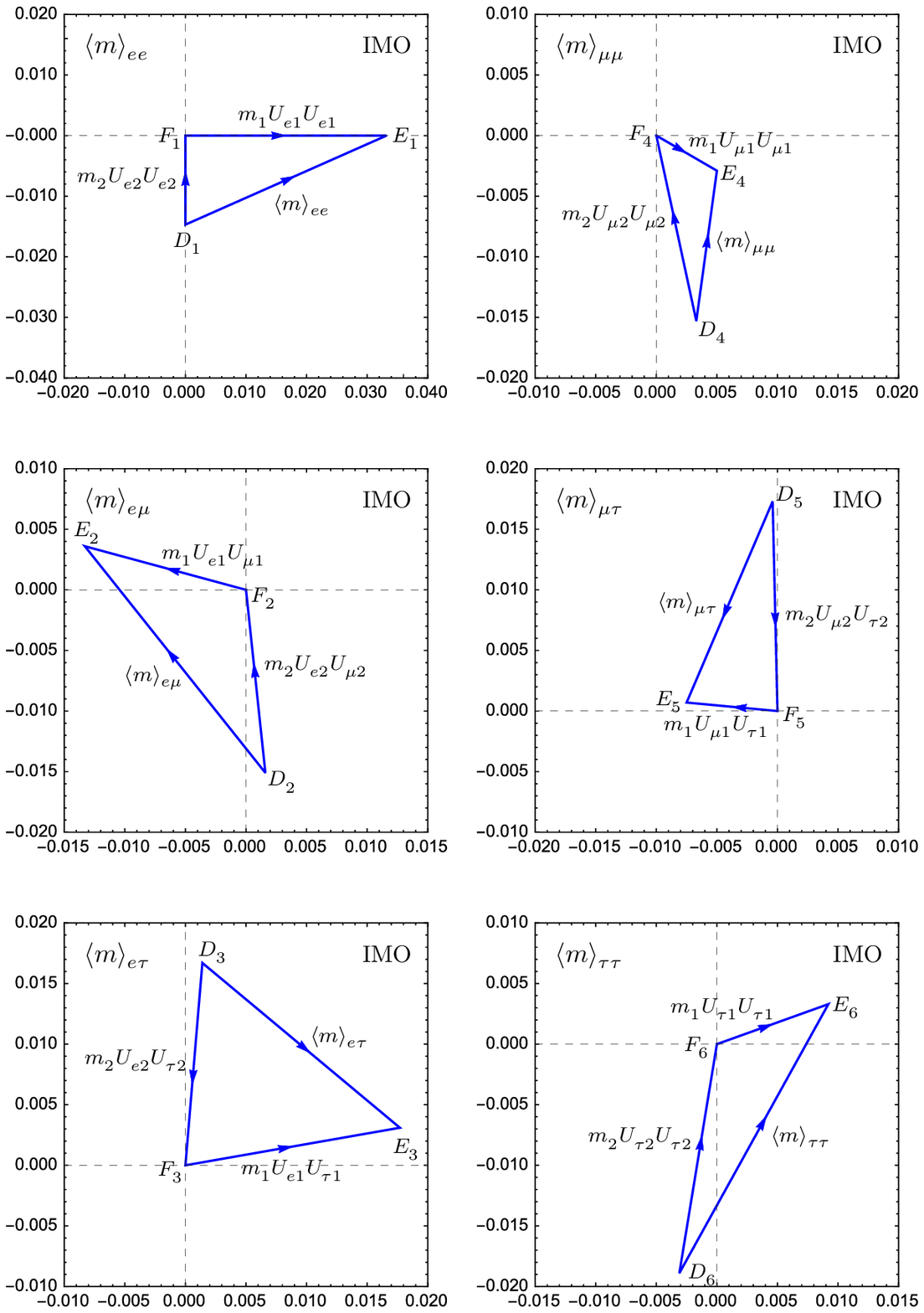}}
\caption{Six effective MTs $\triangle D^{}_i E^{}_i F^{}_i$ (for
$i=1, 2, \cdots, 6$) of the Majorana neutrinos in the $m^{}_3 \to 0$
limit in the complex plane, plotted by assuming the Majorana phase
$\sigma = \pi/4$ and inputting the best-fit values of $\Delta
m^2_{21}$, $\Delta m^2_{32}$, $\theta^{}_{12}$, $\theta^{}_{13}$,
$\theta^{}_{23}$ and $\delta$ \cite{GG} in the IMO case.}
\end{figure}

(1) A remarkable merit of these effective MTs is that they allow us
to easily read off the magnitudes of $\langle
m\rangle^{}_{\alpha\beta}$. For instance, $|\langle
m\rangle^{}_{ee}| \sim |\langle m\rangle^{}_{e\mu}| \sim |\langle
m\rangle^{}_{e\tau}| \sim {\cal O}\left(10^{-3}\right)$ eV and
$|\langle m\rangle^{}_{\mu\mu}| \sim |\langle m\rangle^{}_{\mu\tau}|
\sim |\langle m\rangle^{}_{\tau\tau}| \sim {\cal
O}\left(10^{-2}\right)$ eV in the NMO case; or $|\langle
m\rangle^{}_{\alpha\beta}| \sim {\cal O} \left(10^{-2}\right)$ eV in
the IMO case. Because of $m^{}_2/m^{}_3 \simeq 17.5\%$ in the
$m^{}_1 \to 0$ limit, it is easy to understand why the shortest side
of $\triangle A^{}_4 B^{}_4 C^{}_4$ is $m^{}_2 |U^{}_{\mu 2}|^2$ and
why $|\langle m\rangle^{}_{\mu\mu}| \simeq m^{}_3 |U^{}_{\mu 3}|^2$
holds. The effective MTs $\triangle A^{}_5 B^{}_5 C^{}_5$ and
$\triangle A^{}_6 B^{}_6 C^{}_6$ have a similar property in the NMO
case. In comparison, the $m^{}_3 \to 0$ case is simpler because
$|\langle m\rangle^{}_{\alpha\beta}| \propto m^{}_2 \simeq m^{}_3$
holds.

(2) In the IMO case the effective MT $\triangle D^{}_1 E^{}_1
F^{}_1$ is especially interesting because its inner angle $\angle
D^{}_1 F^{}_1 E^{}_1$ happens to equal $2\sigma$ thanks to $m^{}_3
\to 0$. Therefore, a measurement of $|\langle m\rangle^{}_{ee}|$ of
the $0\nu 2\beta$ decay will allow one to determine the Majorana
phase $\sigma$ in the $m^{}_3 \to 0$ limit. Similarly, $\angle
A^{}_5 C^{}_5 B^{}_5 \simeq 2\sigma$ holds in the $m^{}_1 \to 0$
limit thanks to the smallness of $\theta^{}_{13}$. This observation
implies that it is possible to determine the Majorana phase from a
measurement of the effective mass $|\langle m\rangle^{}_{\mu\tau}|$
in the NMO case with $m^{}_1 \simeq 0$.

(3) The texture of the symmetric Majorana neutrino mass matrix
$M^{}_\nu$, whose six independent elements are just equal to
$\langle m\rangle^{}_{\alpha\beta}$ (for $\alpha, \beta = e, \mu,
\tau$), can be illustrated with the help of Figures 3 and 4 as
follows:
\begin{eqnarray}
|M^{}_\nu| = \begin{pmatrix} |\langle m\rangle^{}_{ee}| &
|\langle m\rangle^{}_{e\mu}| & |\langle m\rangle^{}_{e\tau}| \cr
|\langle m\rangle^{}_{e\mu}| & |\langle m\rangle^{}_{\mu\mu}| &
|\langle m\rangle^{}_{\mu\tau}| \cr
|\langle m\rangle^{}_{e\tau}| & |\langle m\rangle^{}_{\mu\tau}| &
|\langle m\rangle^{}_{\tau\tau}| \cr
\end{pmatrix}
\simeq \displaystyle \left\{
\begin{array}{l}
\begin{pmatrix}
0.0036 & 0.0072 & 0.0033 \cr 0.0072 &
0.0217 & 0.0243 \cr 0.0033 &
0.0243 & 0.0273 \end{pmatrix}
\hspace{0.5cm} (m^{}_1 \to 0) \; , \\
\begin{pmatrix}
0.0363 & 0.0239 & 0.0213 \cr 0.0239 &
0.0125 & 0.0180 \cr 0.0213 & 0.0180 & 0.0254 \cr \end{pmatrix}
\hspace{0.5cm} (m^{}_3 \to 0) \; ,
\end{array}
\right.
\end{eqnarray}
in unit of eV. Such a texture of $M^{}_\nu$ may be reproduced in a
specific neutrino mass model once a kind of flavor symmetry and its
proper breaking are taken into account \cite{Altarelli}.

Note that the probabilities of neutrino-antineutrino oscillations
given in Eq. (13) can be simplified to
\begin{eqnarray}
P\left(\nu^{}_{\alpha} \to \overline{\nu}^{}_{\beta}\right) =
P\left(\overline{\nu}^{}_{\alpha} \to \nu^{}_{\beta}\right) =
\frac{|K|^2}{E^2} \left|\langle m\rangle^{}_{\alpha\beta}\right|^2
\;
\end{eqnarray}
in the $L \to 0$ limit (i.e., the so-called {\it zero-distance
effect}). This result is a clear reflection of the Majorana nature
of massive neutrinos. In fact, the effective Majorana neutrino
masses $\langle m\rangle^{}_{\alpha\beta}$ may also show up in some
other LNV processes, such the $H^{++} \to \alpha^+ \beta^+$ decays
in the type-II seesaw mechanism \cite{SS2}. The branching ratios of
these decay modes are
\begin{eqnarray}
{\cal B}(H^{++} \to \alpha^{+} \beta^{+}) =
\frac{2}{1+\delta_{\alpha\beta}} \cdot \frac{\left| \langle m
\rangle^{}_{\alpha \beta}\right|^2}{\displaystyle m^2_1 + m^2_2 +
m^2_3} \; ,
\end{eqnarray}
where $\alpha$ and $\beta$ run over $e$, $\mu$ and $\tau$. In the
limit of $m^{}_1 \to 0$ or $m^{}_3 \to 0$, one may calculate ${\cal
B}(H^{++} \to \alpha^{+} \beta^{+})$ by inputting the best-fit
values of relevant neutrino oscillation parameters and allowing the
Majorana phase $\sigma$ to vary from $0$ to $2\pi$. The numerical
results are listed in Table 2 for the sake of illustration. Compared
with the previous estimates of such decay modes made some years ago
\cite{Han}, our present results are more convergent because today's
neutrino oscillation data are more accurate and the neutrino mass
limit under consideration is very special. Of course, whether the
type-II seesaw mechanism really works in Nature remains an open
question, and how to measure possible rare LNV processes is a very big
experimental challenge. The point that we are stressing is to see a
potential link between the effective Majorana neutrino masses and
some interesting LNV phenomena.
\begin{table}[t]
\centering \caption{The expected branching ratios of $H^{++} \to
\alpha^+ \beta^+$ decays in the type-II seesaw mechanism, where the
best-fit values of $\Delta m^2_{21}$, $\Delta m^2_{31}$ (or $\Delta
m^2_{32}$), $\theta^{}_{12}$, $\theta^{}_{13}$, $\theta^{}_{23}$ and
$\delta$ \cite{GG} have been input and $\sigma \in [0, 2\pi)$ has been taken.}
\vspace{0.5cm}
\begin{tabular}{|c|c|c|}
\hline Branching ratios & In the $m^{}_1 \to 0$ limit & In the
$m^{}_3 \to 0$ limit \\ \hline ${\cal B}(H^{++} \to e^+ e^+)$ &
$0.0894\% \to 0.5307\%$ & $7.0476\% \to 47.5258\%$
\\
\hline
${\cal B}(H^{++} \to e^+ \mu^+)$ & $0.3426\% \to 4.6215\% $
& $0.0744\% \to 34.8293\%$ \\
\hline
${\cal B}(H^{++} \to e^+ \tau^+)$ & $0.5022\% \to 5.3036\%$
& $0.0613\% \to 50.2589\%$
\\
\hline
${\cal B}(H^{++} \to \mu^+ \mu^+)$ & $14.1339\% \to 24.5069\%$
& $2.0129\% \to 9.4731\%$ \\
\hline
${\cal B}(H^{++} \to \mu^+ \tau^+)$ & $35.0848\% \to 58.3642\%$
&  $3.9764\% \to 25.5264\%$
\\
\hline
${\cal B}(H^{++} \to \tau^+ \tau^+)$ & $21.7296\% \to 34.7906\%$
& $1.8257\% \to 17.3884\%$ \\
\hline
\end{tabular}
\end{table}

Assuming that a positive signal of the $0\nu 2\beta$ decay can be
measured someday, then the corresponding knowledge of $|\langle
m\rangle^{}_{ee}|$ will allow one to predict the rates of some other
rare LNV processes, such as $P(\nu^{}_e \to \overline{\nu}^{}_e) =
|K|^2 |\langle m\rangle^{}_{ee}|^2/E^2$ from Eq. (21) and ${\cal
B}(H^{++} \to e^{+} e^{+}) = |\langle m \rangle^{}_{ee}|^2/
\left(m^2_1 + m^2_2 + m^2_3\right)$ from Eq. (22). Once such a
breakthrough really happens, it will definitely open a new window
towards the deep secrets of Majorana particles.

\section{Summary}

Neutrino physics has entered the era of precision measurements, in
which one is doing the best one can to answer some important
questions, including what the absolute neutrino mass scale is,
whether massive neutrinos are the Majorana particles, how large the
effects of leptonic CP violation can be, and so on. Before these
questions are experimentally answered, one may theoretically or
phenomenologically try every shift available to bridge the gap
between the observable quantities and the fundamental flavor
parameters in the neutrino sector. In this regard we have paid
particular attention to an intuitive description of leptonic CP
violation and effective Majorana neutrino masses in the complex
plane --- namely, the Dirac and Majorana UTs as well as the
effective MTs in the $m^{}_1 \to 0$ or $m^{}_3 \to 0$ limit.

With the help of the best-fit values of neutrino oscillation
parameters, we have plotted the six UTs of the PMNS matrix to show
their real shapes in the complex plane. The connections of the
Majorana UTs with neutrino-antineutrino oscillations and neutrino
decays have been explored, and the possibilities of right or
isosceles UTs have also been discussed. In the second part of this
paper, we have considered a special but phenomenologically allowed
neutrino mass spectrum with $m^{}_1 =0$ or $m^{}_3 =0$ and the
corresponding effective Majorana neutrino masses $\langle
m\rangle^{}_{\alpha\beta}$ --- the latter can form six MTs in the
complex plane. In this case we have shown how these MTs look like by
assuming the Majorana phase $\sigma$ to be $\pi/4$ as a typical
example. The relations of such triangles to the LNV decays $H^{++}
\to \alpha^+ \beta^+$ in the type-II seesaw mechanism have been
illustrated too.

We hope that this kind of study may enrich the neutrino
phenomenology to some extent. Although the UTs and MTs can only
provide us with a geometrical language to describe the flavor issues
of massive neutrinos, they {\it do} have made some underlying
physics more transparent and intuitive. So they are useful and
interesting, and their phenomenological applications deserve some
further exploration.


This work was supported in part by the National Natural Science
Foundation of China under grant No. 11135009 and No. 11375207.


\end{document}